# A network-based biomarkers discovery of Cold/Hot ZHENG chronic gastritis and Cold/Hot herbs of formulae

Boyang Wang[a], Pan Chen[a], Peng Zhang[a] and Shao Li[a,*]

[a]Institute for TCM-X, MOE Key Laboratory of Bioinformatics, Bioinformatics Division, BNRist, Department of Automation, Tsinghua University, 100084 Beijing, China

## Abstract

Objective: To discover biomarkers and uncover the mechanism of Cold/Hot ZHENG (syndrome in traditional Chinese medicine) chronic gastritis (CG) and Cold/Hot herbs in traditional Chinese medicine (TCM) formulae on systematic biology.

Background: CG is a common inflammatory disease and the diagnosis of CG in TCM can be classified into Cold ZHENG (Asthenic Cold) and Hot ZHENG (Excess Hot). However, the molecular features of Cold/Hot ZHENG in CG and the mechanism of Cold/Hot herbs in formulae for CG remained unclear.

Methods: Based on data of 35 patients of Cold/Hot ZHENG CG and 3 scRNA-seq CG samples, we conduct analysis with transcriptomics datasets and algorithms, to discover biomarkers for Cold/Hot ZHENG CG. And we collected 25 formulae (with traditional effects related to Cold/Hot ZHENG) for CG and corresponding 89 Cold/Hot herbs (including Warm/Cool herbs) to discover features and construct target networks of Cold/Hot herbs on the basis of network target and enrichment analysis.

Results: Biomarkers of Cold/Hot ZHENG CG represented by CCL2 and LEP suggested that Hot ZHENG CG might be characterized by over-inflammation and exuberant metabolism, and Cold ZHENG CG showed a trend of suppression in immune regulation and energy metabolism. And biomarkers of Cold/Hot ZHENG showed also significant changes in the progression of gastric cancer. And biomarkers and pathways of Hot herbs intend to regulate immune responses and energy metabolism, while those of Cold herbs were likely to participate in anti-inflammation effect.

Conclusion: In this study, we found that the biomarkers and mechanism of Cold/Hot ZHENG CG and those of Cold/Hot herbs were closely related to the regulation of immune and metabolisms. These findings may reflect the mechanism, build bridges between multiple views of Cold/Hot ZHENG and Cold/Hot herbs, and provide a research paradigm for further achieving precision TCM.

**Keywords**: Cold/Hot ZHENG, chronic gastritis, Cold/Hot properties, traditional Chinese medicine, network targets

## 1. Introduction

Chronic gastritis (CG) is defined as inflammatory diseases of the gastric mucosa and is classified as chronic superficial gastritis (CSG) and chronic atrophic gastritis (CAG) based on the histopathologic patterns and endoscopic appearances of the gastric mucosa [1]. The prevalence of CG may exceed 50% worldwide [2]. The progressive deterioration of atrophic gastritis, which subsequently leads to dysfunction of the gastric mucosa, is also the highest independent risk factor for gastric cancer [3].

To date, the etiology of CG remains incompletely understood. It can be caused by a range of factors such as stress, alcohol, irrational use of nonsteroidal anti-inflammatory drugs, and *H. pylori infection*, leading to an imbalance between offensive acid-pepsin secretion and defensive mucosal factors such as cell shedding and mucin secretion [4]. Regarding stress risks, physiologic stress can result in dysregulation of gastric pH, which contributed to gastritis. In the stressed state, increased levels of histamine and acetylcholine result in elevated acid production, thus inducing or worsening gastritis [5]. The present therapy of gastritis is to alleviate inflammation and associated dyspeptic symptoms, and specific treatments should be determined accordingly depending on each individual's condition. In China, traditional Chinese medicine (TCM) therapy is an important complementary treatment option for CG [6].

ZHENG, a TCM theoretical understanding of the symptomatic profiles of disease, is used to recognize and understand the non-healthy physiological states of patients from a holistic view. In TCM clinical diagnosis, there are different types of ZHENG for the same disease depending on the phenotype profiles. All diagnostic and therapeutic approaches in TCM are based on the typology of ZHENG. In the diagnosis of gastritis in TCM, the patients with CG can be classified into two main types: Cold ZHENG and Hot ZHENG. The CG associated with Cold ZHENG has characteristics of cooling of limbs, loose stool, clear abundant urine, white-greasy tongue coating. The CG associated with Hot ZHENG is characterized by red tongue, yellow-dense tongue coating, thirst, dry mouth, deep-colored urine, and dysphoria with a feverish sensation. In our previous study, we have evaluated the biological basis of CG associated with Cold/Hot ZHENG, suggesting that the metabolism-immune network imbalance has the potential to be a new perspective in the development of sub-typing and individualized treatment for CG [7]. According to the rules of TCM diagnosis and treatment, patients with Cold ZHENG should be treated by herbs with hot property (Hot herbs) and patients with Hot ZHENG should be treated by herbs with cold property (Cold herbs) for thousands of years in China. For the treatment of CG in TCM, CG associated with Cold and Hot ZHENG are treated with herbs that have hot and cold properties, respectively. However, the biological mechanisms behind Hot and Cold herbs for the treatment of Cold and Hot ZHENG remain unclear. Recent advances in TCM research are currently associated with the rapid development of concepts in network pharmacology and systems biology that provide approaches to understanding the rules of TCM diagnosis and treatment [8-10]. The aim of this study is to investigate the mechanisms of action of Cold/Hot herbs in the treatment of CG associated with Cold/Hot ZHENG through a network pharmacology approach.

## 2. Results

### 2.1 Outlier of the whole study

As shown in Figure1, we conducted a comprehensive analysis on gastritis and widely used TCM for gastritis from the perspective of Cold/Hot ZHENG and network targets in this study. First and foremost, based on seeds genes from Cold/Hot biological network [11] proposed by Li and microarray for CAG and CSG with Cold/Hot ZHENG [7], we constructed Cold/Hot biological network for CG to discover the features and biomarkers of Cold/Hot ZHENG. On the basis of the biological network and machine learning algorithms, features and biomarkers of immune and metabolism were obtained.

Besides, we collected 29 formulae for GC and corresponding 132 herbs recorded in these formulae from Pharmacopoeia of China. Cold/Hot information and Meridian information of these herbs were collected from Pharmacopoeia. The distribution of Meridian information of the herbs were counted and the compounds composition of these herbs was obtained from commonly-used TCM database Herbiomap [12] and Symmap [13]. After filtering herbs without recording or without, 25 formulae, 89 Cold/Hot herbs, 19 other herbs and 2853 compounds were kept for further study and the targets profiles for these herbs and formulae were characterized by our previous network-based algorithms [14,15]. Based on the constructed network, features and biomarkers of Cold/Hot which were mainly related to immune and metabolism were acquired.

Finally, as a combination of Cold/Hot biological network for CG and targets profiles for Cold/Hot TCM, Cold/Hot targets biological network of Cold/Hot TCM for CG. This network described the most-frequently targeted Cold/Hot genes of Cold/Hot TCM and uncovered the mechanism of Cold/Hot CG and corresponding Cold/Hot herbs in formulae for CG to some extension.

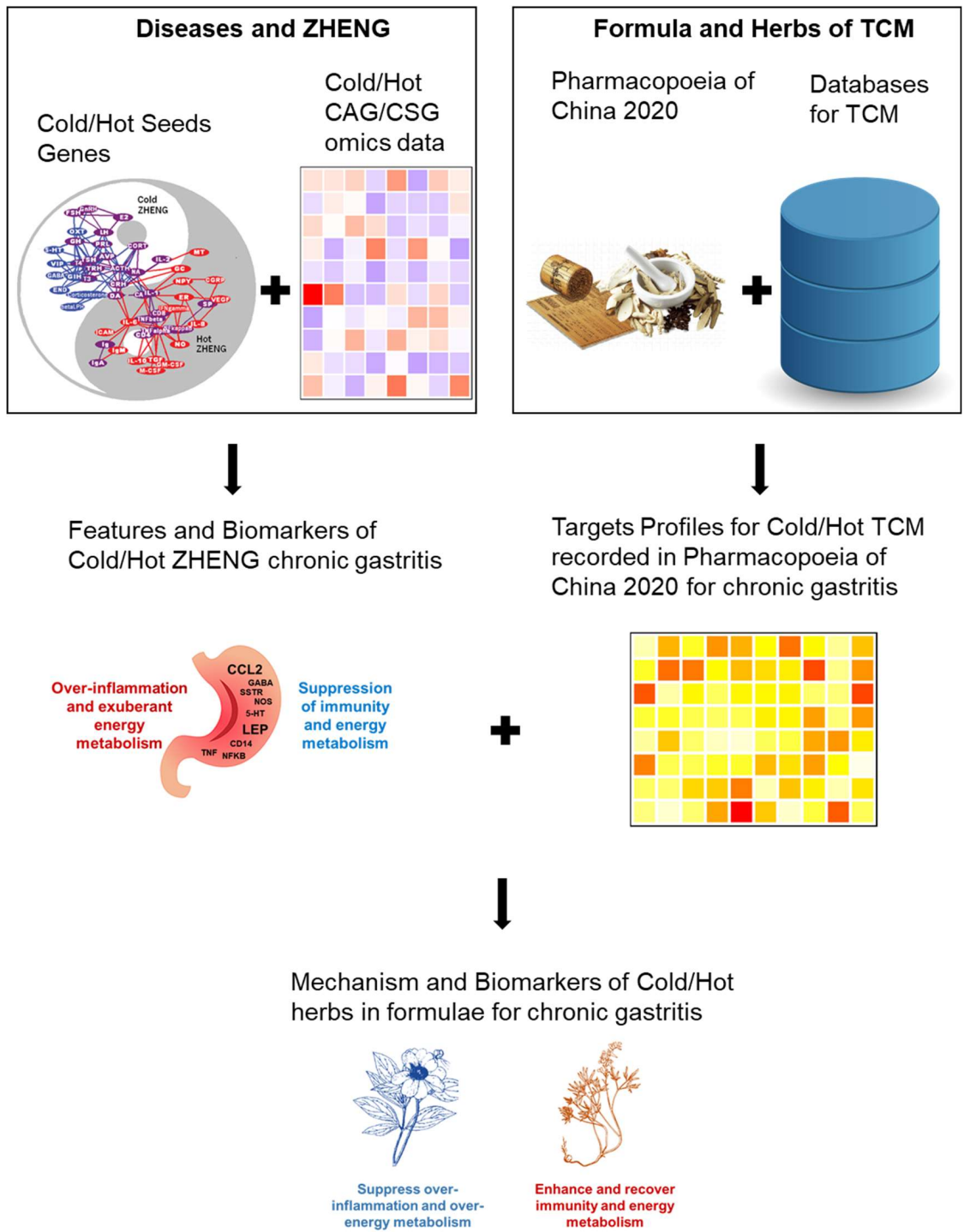

**Figure 1.** The overall outlier of the study from two perspectives including Cold/Hot ZHENG CG and Cold/Hot herbs

## 2.2 Molecular features of Cold/Hot ZHENG CG

In 2013, Li collected 35 patients of Cold/Hot ZHENG CG for microarray measurement, including 17 patients with Cold ZHENG (8 for chronic superficial gastritis and 9 for chronic atrophic gastritis) and 18 patients with Hot ZHENG (8 for chronic superficial gastritis and 10 for chronic atrophic gastritis). In order to find key molecules related to Cold/Hot ZHENG in CG, we collected seeds gene from previous Cold/Hot network model as background for Cold/Hot ZHENG. PLS-DA analysis successfully grouped Hot and Cold ZHENG CG for CAG patients and CSG patients, respectively (Figure 2A). VIP (variable importance) for each gene in CAG patients or CSG patients was calculated and 26 of the seeds genes with VIP greater than 1 both occurred in CAG patients and CSG patients (Figure 2B).

Besides, from another perspective, DEGs (differentially expressed genes) in CAG patients and CSG patients were calculated by limma model to find significantly expressed genes between Cold/Hot ZHENG in both CAG patients and CSG patients. Among these DEGs, 112 genes were differentially expressed in both CAG patients and CSG patients (adjust p value <0.05, BH adjustment). And 11 genes were up-regulated in both Cold ZHENG CAG patients and CSG patients, while 47 of them were up-regulated in both Hot ZHENG of two disease conditions (Figure 2C). Combining these two analyses of different perspectives, both in statistical methods and hierarchical clustering methods, Hot ZHENG patients were distinct from Cold ZHENG patients (Figure 2D). This result suggested that there existed gene expression patterns between these two conditions of CG which might be able to mined from our analysis and further constructed biological networks.

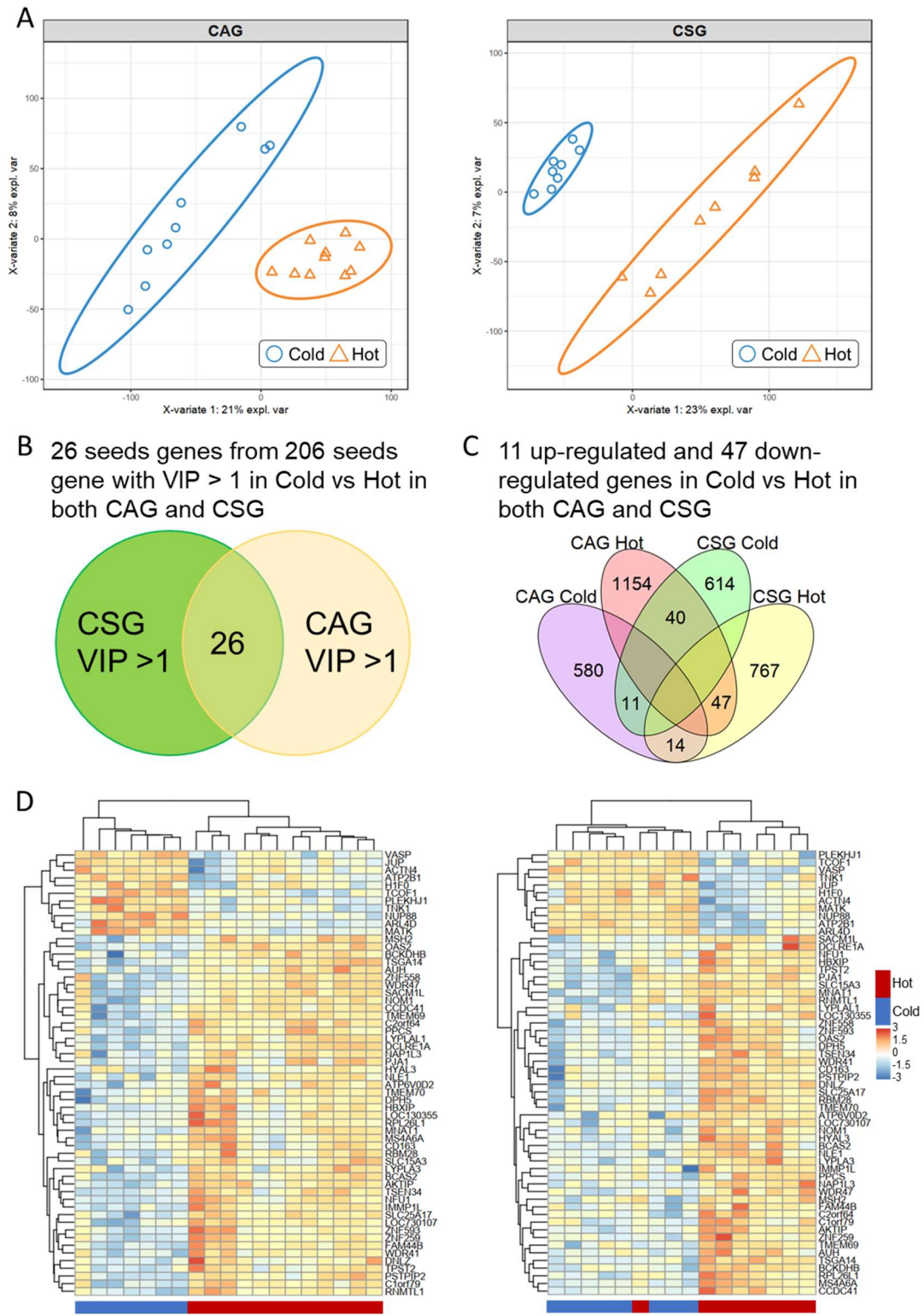

**Figure 2.** Analysis for finding representative molecules between Cold/Hot ZHENG CG. (A) PLS-DA analysis for Cold/Hot ZHENG CAG (left) and CSG (right). (B) Venn plot showing genes with VIP larger than 1 in both Cold/Hot ZHENG CAG and CSG. (C) Venn plot showing the differential expression of genes in microarray of Cold/Hot ZHENG CAG and CSG. (D) Heat map of 47 genes

both up-regulated in Hot ZHENG CAG and CSG and 11 genes up-regulated in Cold ZHENG CAG and CSG for CAG patients (left) and CSG patients (right).

**2.3 Immune and metabolic characteristics of Cold/Hot ZHENG CG**

Based on our found molecular features of Cold/Hot ZHENG CG, we further payed attention to the biological processes or pathways enriched by these molecular features. Firstly, we performed KEGG pathway enrichment [16] and Gene Ontology (GO) enrichment [17] on 1846 DEGs (1241 genes up-regulated in Cold ZHENG and 605 genes up-regulated in Hot ZHENG) of CAG patients. It was found that pathways and biological processes related to immune and metabolism were significantly enriched (Figure 3A). Further, we performed Gene Set Enrichment Analysis (GSEA) [18] on CAG patient and GSEA terms that significantly enriched were shown in Figure x. It could be found that, GSEA terms related to immune, inflammation, cytokines and chemokines (Figure 3B-D) were activated in Hot ZHENG CAG patient (also can be defined as inhibited terms in Cold ZHENG CAG patient, NES < -1), while terms related to metabolism and secretion of peptide, hormone, steroid (Figure 3B, E), as well as cellular junctions and adhesion were activated in Cold ZHENG CAG patient (NES > 1). These findings suggested that in Hot ZHENG CG, pathways or biological processes related to immune and inflammation might be over-developed, while in Cold ZHENG CG, the main distinguished features turned out to be activating in endocrine and energy metabolism. Apart from CG, these modules of immune regulation and metabolism had also been reported in researches about other diseases [19-21].

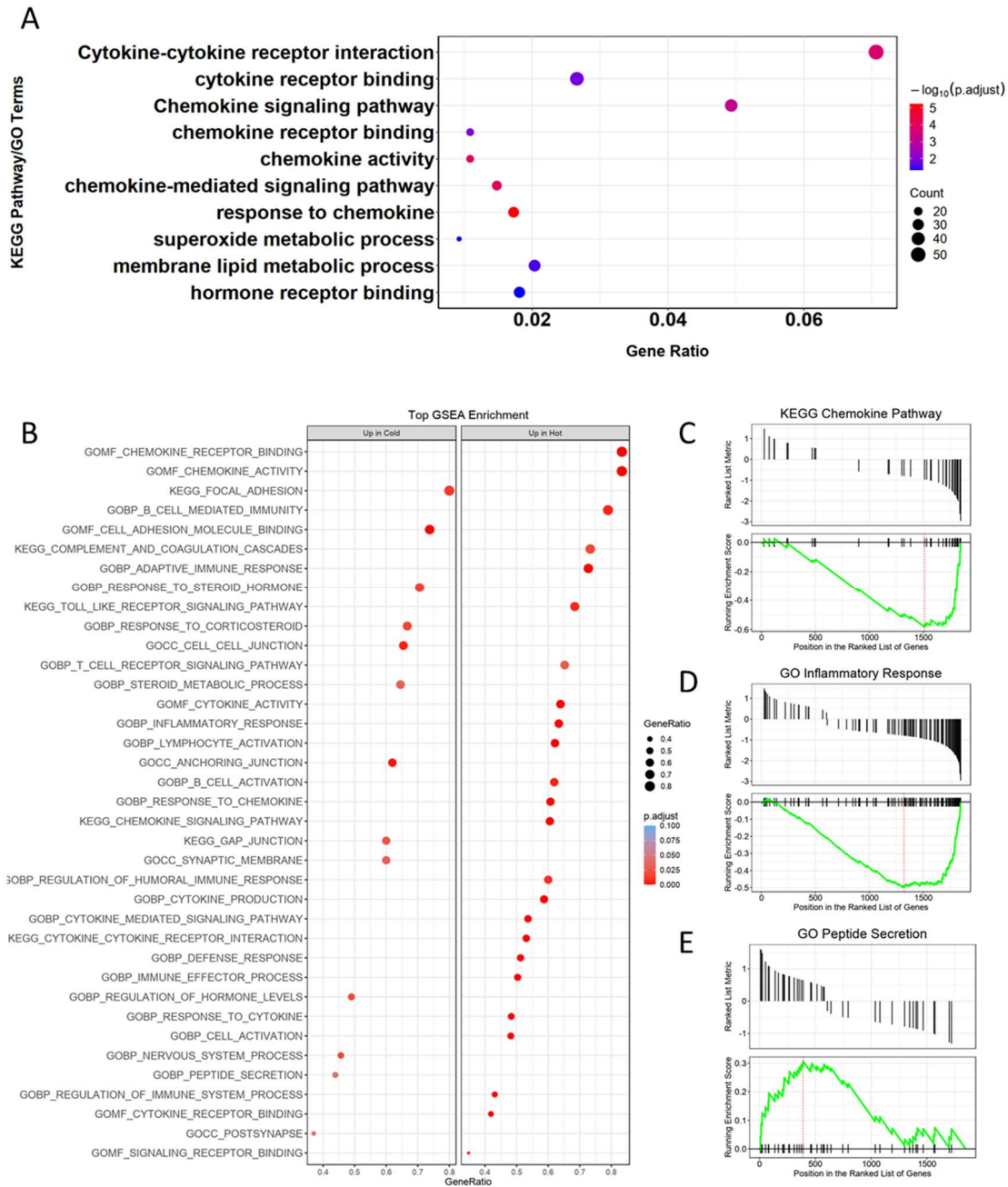

**Figure 3.** Enrichment analysis for immune and metabolic characteristics of Cold/Hot ZHENG CG (A) Dot plot of KEGG and GO enrichment analysis for 1846 DEGs of CAG patients. (B) GSEA for genes and their expression of CAG patients. (C)-(D) GSEA result for Chemokine Pathway, Inflammatory Response and Peptide Secretion, respectively.

As inferred by our previous findings, we focused on the immune and inflammation characteristics in CG. Based on CIBERSORT algorithm [22], proportions of different immune cells were deconvoluted in Cold/Hot ZHENG CG, respectively. It could be found that some immune cells, represented by M1 macrophages, showed a significantly different proportion in Hot ZHENG CG than Cold ZHENG CG (Figure 4A). The proportions of both M1 and M2 macrophages were significantly higher in Hot ZHENG CG than that of Cold ZHENG CG, which might confirm the findings that from the perspective of immune and inflammation regulation, the most distinguishing features between Hot ZHENG and Cold ZHENG gastritis is the over-inflammation in Hot ZHENG and the suppression of immune in Cold ZHENG.

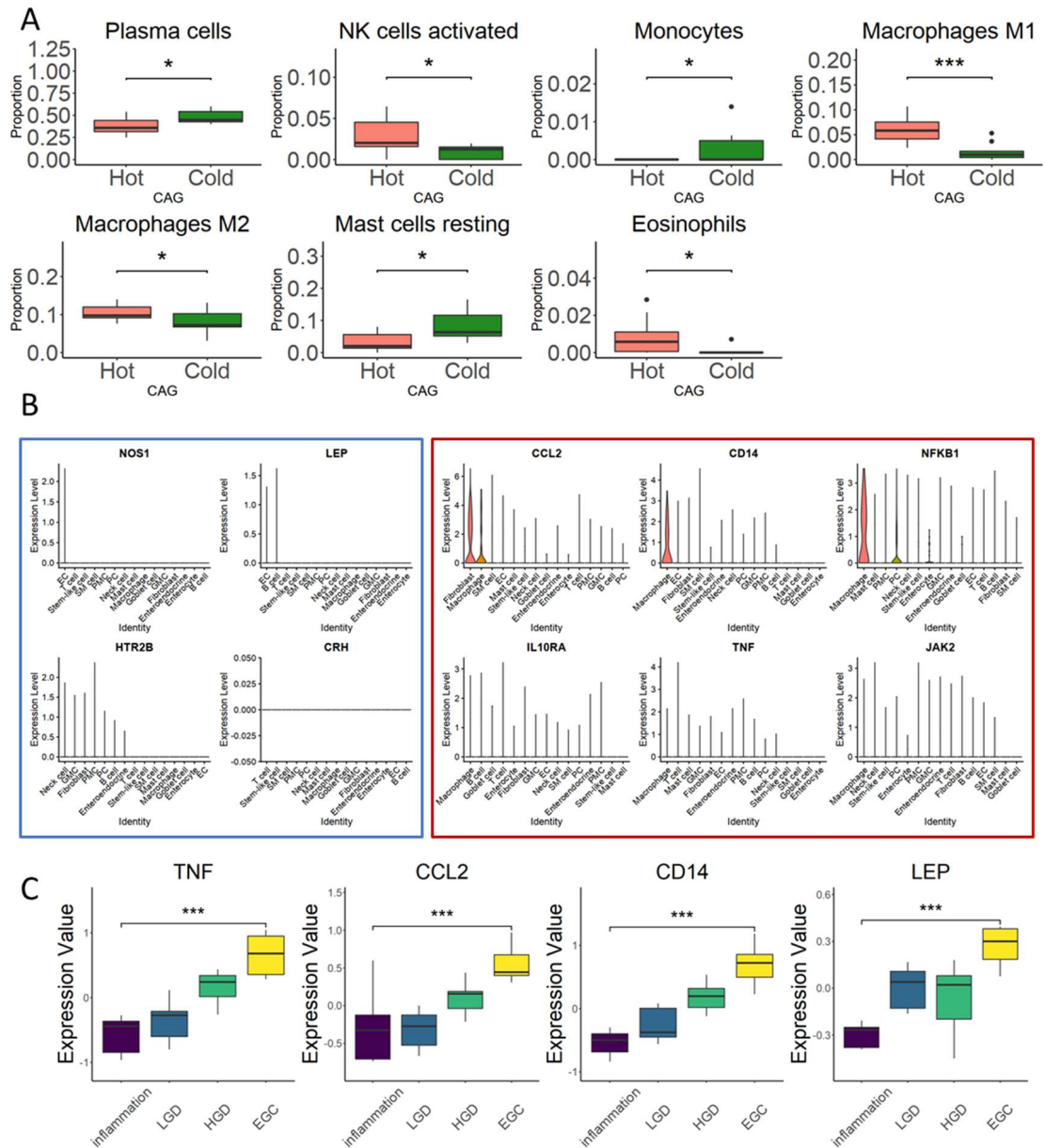

**Figure 4.** Network construction of Cold/Hot ZHENG CG. (A) Inference of the proportion of immune cells significantly changed in Cold ZHENG CAG and Hot ZHENG CAG. (B) Expression of previous reported and newly found biomolecules for Cold/Hot ZHENG CAG in single-cell level of CAG patients. (C) Box plots showing the expression of biomolecules during the progression of gastric cancer.

In addition, we filter out 3 Hot ZHENG CAG samples from large scale scRNA-seq of human gastric cancer progression [3]. In this cellular level measurements of expression of genes in these characteristic pathways and biological processes related to immune regulation and metabolism, we focused on some key molecules in Cold/Hot ZHENG CG. It was found that biomarkers of Cold ZHENG CG which participated in these key pathways and biological processes, like HTR2B, CRH, NOS1 and LEP, hardly expressed in any cell types in Hot ZHENG CG samples. These genes were reported in previous study [7], and were found to be related to key link in Cold ZHENG, including 5-HT related gene HTR2B, corticotropin releasing hormone related genes CRH, CRHR1 and POMC, leptin related gene LEP and nitric oxide related gene NOS1. On the contrary, genes related to immune and inflammation were relatively higher expressed, especially in macrophages, which was consistent with our above findings that macrophages significantly increased in Hot ZHENG CG. These genes included CCL2, CD14, NFKB1, IL10RA, TNF and JAK2, which were related to inflammation, cytokines, chemokines and immune regulation (Figure 4B). Based on public omics data, it was also found that some of these biomarkers showed significant changes in the progression from gastritis to dysplasia and gastric cancer (Figure 4C), and were reported to participant in the progression of gastric cancer, such as LEP, CCL2, CD14 and TNF. The expression of these biomolecules and their related biomolecules was found to be associated with gastric cancer progression and prognosis [23-26], which also inferred that pathways or cells related to these biomarkers may also play roles in cancer progression and prognosis. Besides, some other seeds genes also showed a significantly differential expression in the progression of gastric cancer (Figure S2). These findings inferred us that the key biomolecules in Cold/Hot ZHENG might also play important roles in other disease progressions which need further analyses and researches. These complex features in immune regulation and metabolism, as well as biomolecules like TNF, VEGF, TGFB and NFKB1, also reflect the potential risk of CG, especially CAG in inflammation-induced tumorigenesis according to our previous constructed tumorigenesis network [27,28], and also be reported in researches of tumorigenesis in other digestive systems diseases like chronic hepatitis and enteritis [29,30].

Finally, based on combination of multi-omics data and machine learning algorithms, we constructed a homogeneous biological network, composed of key seeds genes of Cold/Hot ZHENG and DEGs in both two kinds of CG (Figure S1). In this network, the interactions between every two genes were collected from STRING database. It could be found that many of the seeds genes played important roles in these network with high connections, especially immune and inflammation related genes like CCL2, CD14, NFKB1, IL2RB, JAK2, VEGFC, TGFB3 and IL10RA, most of which showed high expression in macrophages of Hot ZHENG CG patients. Besides, some genes related to endocrine and energy metabolism including SSTR2, SSTR5, HTR1A, CRH, CRHR1 and POMC also had high degrees in this network. It is worthy to be noticed that not only the biomolecules in the network for Cold/Hot ZHENG CG, but also the genes with close functional relationship or biological relationship may play important role in the further diagnosis and mechanism uncovering of Cold/Hot ZHENG CG.

According to our enrichment result, metabolism features of CG is also of vital importance in the diagnosis of Cold/Hot ZHENG CG. Metabolism related to peptide is significantly enriched, and the interactions between peptide and protein participate in various fundamental cellular functions [31]. The pathologically elevated steroid hormones may be accompanied by leptin resistance, which weakens normal energy expenditure and thermogenesis [32]. In our previous study, we found that serum level of leptin in CAG patients associated with Cold ZHENG was significantly higher than normal subjects [7]. Therefore, the presence of pathologically elevated leptin levels in patients with cold ZHENG means that their reduced energy expenditure and thermogenesis may be due to leptin resistance. Conditional Dlx1/2-null mice showed a loss of growth hormone-releasing hormone neurons with higher somatostatin expression and lower energy expenditure [33]. A previous study also showed that somatostatin in the paraventricular nucleus of the hypothalamus could inhibit thermogenesis [34], suggesting that SSTR involves in energy expenditure. It has been reported that 5-HT could inhibit thermogenesis through Htr3 in brown adipose tissue

[35]. Besides, in the median preoptic nucleus, the thermoregulatory response is initiated by stimulation of GABA neurons, suggesting that GABA plays an important role in the process of immune regulation and energy expenditure [36]. Last but not the least, the suppression of tight junction and gap junction was associated with the activation of gene networks of adaptive immunity [37]. And tight junction was reported to be related to immune suppression in COVID-19 [38].

**2.4 Characteristics of formulae for CG**

From the perspective of system biology, we focused on the potential effect on formulae for CG. We measured some typical pathways or biological processes for Cold/Hot ZHENG CG in our above findings in all 29 formulae recorded in Pharmacopoeia. It was found that in specific pathways or biological processes in immune regulation, inflammation and steroid dominated energy metabolism, the potential effect of different formulae differed from others in immune-related pathways and biological processes like immune cells, immune response, cytokines and chemokine. Besides, some other pathways and biological processes related to energy reverse metabolism, nitric oxide. On the contrary, steroid metabolic process, response to steroid hormone, steroid hormone mediated signaling pathway, regulation of inflammatory response and response to oxidative stress were consistently significantly enriched and might be a coincident potential mechanism of these formulae against CG (Figure 5A).

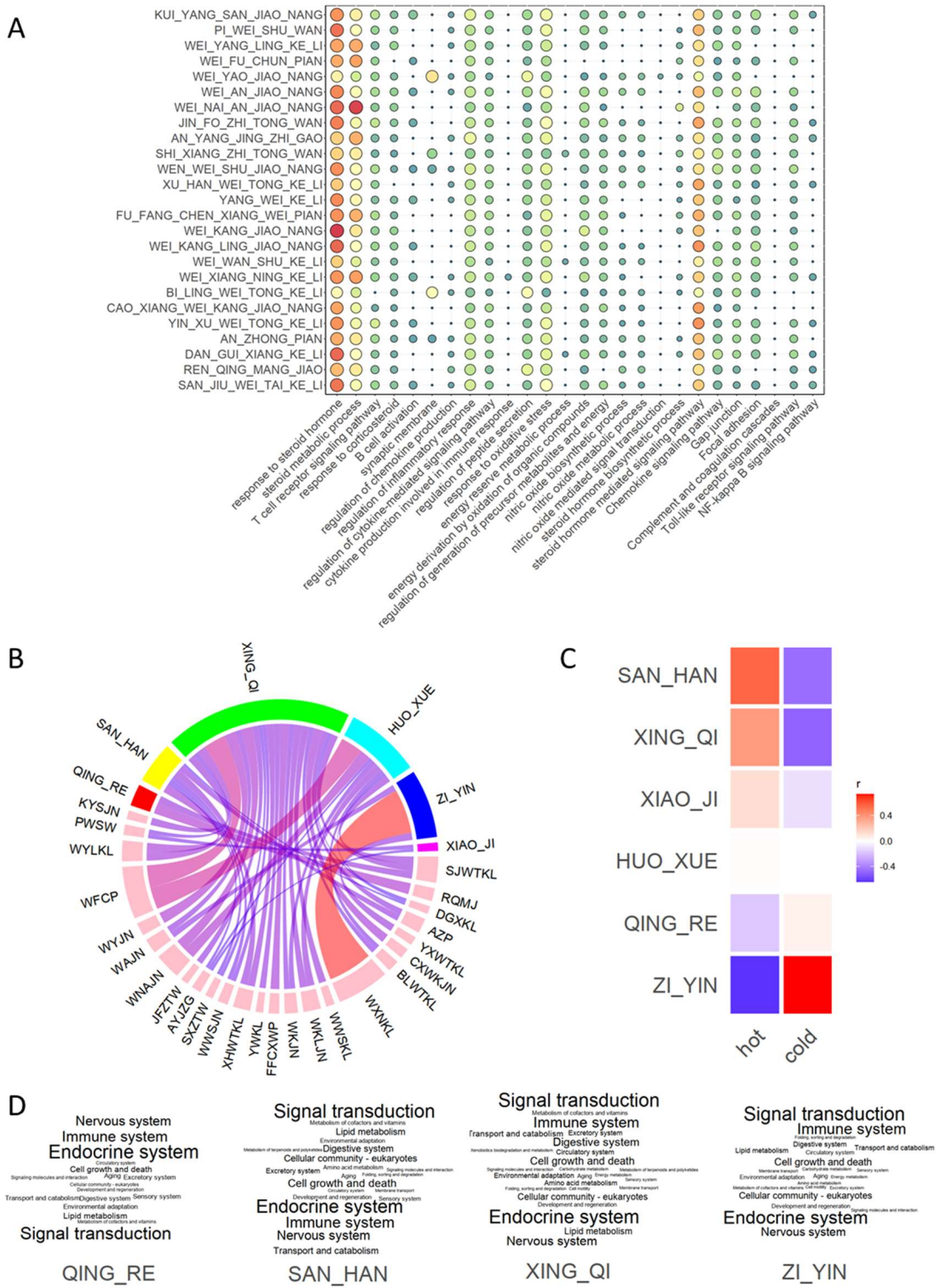

**Figure 5.** Potential mechanism of formulae recorded in Pharmacopoeia of China for CG. (A) Dot plot showing the potential effect of formulae on the representative pathways and biological processes of Cold/Hot ZHENG CG. (B) Occurrence of the six labels including ZI YIN, XIAO JI, SAN HAN, QING RE, HUO XUE and XING QI in these formulae. (C) Heat map showing the correlation between six labels and the proportion of Cold or Hot herbs in a formula. (D) Word cloud

showing pathways enriched by formulae with different traditional effects.

All the 25 formulae could be pasted with six labels for traditional effects, including ZI YIN, XIAO JI, SAN HAN, QING RE, HUO XUE and XING QI. According to TCM experience, these six labels were corresponded with specific effects, in which ZI YIN, XIAO JI, SAN HAN, QING RE, HUO XUE and XING QI means nourishing humors, eliminating food stagnation, dispelling Cold, removing Hot, promoting the restoration of vitality and achieving smooth air flow, respectively. As shown in Figure 5B and 5C, XING QI was the most frequent effect for these formulae, and it was positively correlated (Wilcoxon test, P value < 0.05) with the proportion of Hot TCM in a formula. Besides, ZI YIN had significantly positive correlation with the proportion of Cold TCM, while SAN HAN had significantly positive correlation with the proportion of Hot TCM. In addition, XIAO JI and XING QI showed a relatively positively correlation with the proportion of Hot TCM in a formula. These findings might uncover the material basis of these related six labels for traditional effects. Further, in the four significantly different labels for traditional effects in formulae with Cold/Hot herbs, we focused on the pathways of formulae with them to uncover the mechanism of these four traditional effects. It was found that pathways belonging to signal transduction, endocrine and immune were the most important in these four traditional effects, which might consistently support our findings that metabolism and immune regulation were the key mechanism of Cold/Hot ZHENG CG.

### 2.5 Herbs and depiction of their target profiles in formulae against CG

Another main finding of this study falls in the various mechanism of actions of Cold/Hot herbs in traditional formulae for CG. We collected 29 traditional formulae for CG and their corresponding herbs from Pharmacopoeia. These 29 formulae totally included 242 herbs (132 unique herbs). In annotating the compound composition for these herbs from two large databases Symmap and Herbiomap, 108 of these 132 herbs were successfully matched with the compounds as well as Cold/Hot information and meridian information. In total, 2853 unique compounds were found in these herbs and annotated with PubChem CID [39] for further targets prediction.

Target profiles of these compounds were calculated by our previous network-based algorithm DrugCIPHER-SC [14] and top 100 druggable targets of the profiles were chosen as the targets for further analysis. In order to measure the holistic targets of formulae and herbs, a previous statistical strategy [15] was performed and targets with occurrence significance less than 0.05 (BH adjustment) were chosen.

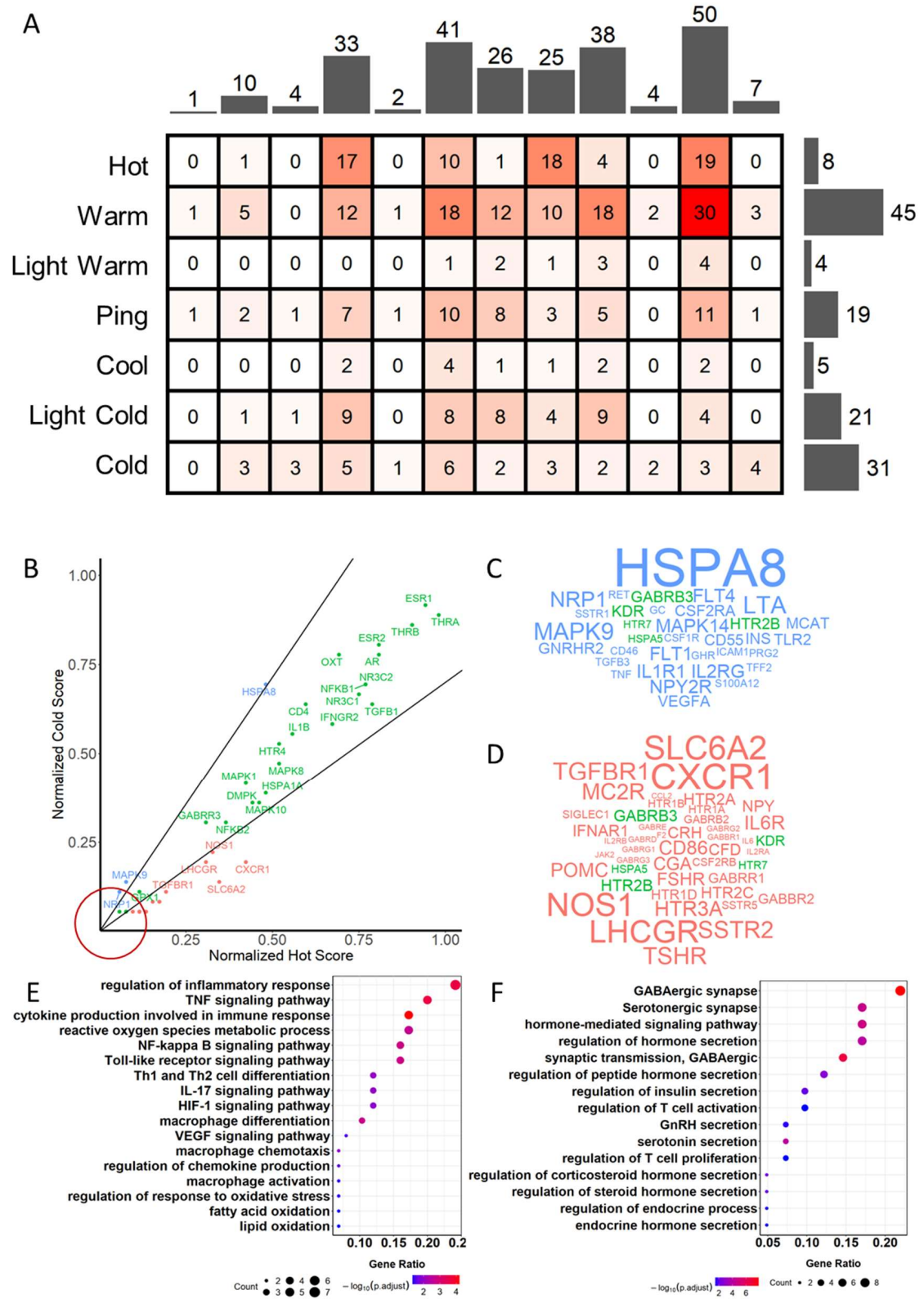

**Figure 6.** Analysis of Cold/Hot herbs in formulae for CG. (A) Heat map showing the Cold/Hot properties and meridian of these Cold/Hot herbs. (B) Visualization of the normalized Cold/Hot score of Cold/Hot ZHENG seeds genes in which some of the genes near two axes weren't shown and were hidden in the read circle. (C)-(D) Wordcloud plot for targets of Cold/Hot herbs and shared

targets in which word size represents the ratio of normalized Cold score and normalized Hot score. (E)-(F) Dot plot showing KEGG and GO enrichment of targets of Cold/Hot herbs.

Cold/Hot information is one of the most important information in TCM. According to Pharmacopoeia, the Hot and Cold properties have been divided into seven levels: Cold, light Cold, Cool, Ping (a kind of state which has no tendency to Cold or Hot), light Warm, Warm and Hot (including great Hot). The first three levels belong to Cold category and the last three levels belong to Hot category. Herbs belonging to Warm held the largest population with the amount of 45 and those of Cold held the second largest population of 31, while herbs belonging to Hot, light Warm and Cool held the least population.

Apart from Cold/Hot information, meridian was also of vital importance for herbs, for the reason that it might figure out the place where herbs took effect. Herbs in traditional formulae for CG mainly includes 12 kinds of meridian, including triple energizer (tri-jiao, a special TCM term), large intestine, small intestine, heart, pericardium, liver, lung, kidney, stomach, gallbladder, spleen and bladder. It was found that one herb might include more than one meridian, which showed that herbs might take effect in many tissues. It could be found that spleen was the most frequent destination of these TCM and liver, stomach, heart, lung and kidney ranked the second to the sixth. On the contrary, triple energizer, pericardium, gallbladder and bladder ranked at the end with the amount less than 10. Besides, we also observed the cross mapping of Cold/Hot properties and meridian (Figure 6A), and it was worthy to be noticed that the most frequent pair was Warm and spleen. This finding was corresponding with previous hypothesis, for the reason that spleen is an important immune-related organ and one of the most important effect of Hot TCM is enhancing immune regulation. However, the distribution of Cold TCM seems to be more concentrated in stomach, kidney, liver, lung and heart rather than spleen.

## 2.6 Molecular features of targets of Cold/Hot TCM for CG

In consideration of the complex composition of TCM, many herbs might share the same targets according to network target analysis. So, in order to find the potential tendencies for Cold/Hot TCM respectively, we set a threshold of 0.7 (see Methods) and divided targets of Cold/Hot TCM into three classes: Hot TCM targets, Cold TCM targets and shared targets. Shared targets were the class of targets that might be targeted by a number of Hot TCM and Cold TCM at the meantime (Figure 6B). Biomolecules related to inflammation and immune regulation like TNF, ILR1, VEGFA, TLR2 and IL2RG were classified as Cold TCM targets, inferring that potential effects of Cold TCM might including anti-inflammation and immune regulation (Figure 6C). On the other hand, those involved in energy metabolism including metabolic process of steroid, hormone, 5-HT, SSTR, NO, CRH and GABA were targeted by Hot TCM, together with those playing import roles in immune response like IL6R, CXCR1 (Figure 6D). Besides, shared targets were composed of biomolecules participating many important biological processes, like IL1B, CD4, AR, ESR1, ESR2, NFKB1, TGFB1.

KEGG and GO enrichment were also performed on Hot herbs targets and Cold herbs targets. It was found that, in the enrichment result of Cold herbs targets, biological processes related to inflammatory response, cytokines, chemokines and immune cells represented by macrophages were significantly enriched, as well as inflammatory pathways including TNF, VEGF, HIF-1a signaling pathways. Apart from these immune-related pathways or biological processes, those related to lipid metabolism were also significantly enriched, like fatty acid oxidation and lipid oxidation (Figure 6E). The enrichment result of Hot herbs targets was much different, which mainly fell in biological processes bound up with inhibitory neurotransmitter like 5-HT, GABA, hormone including steroid hormone, corticosteroid hormone, peptide hormone and other endocrine hormones (Figure 6F). Besides, the result also included cellular processes of T cells, like T cell activation and proliferation. In general, these findings showed that in the one hand, the mechanism of Cold herbs against Hot ZHENG mainly fell in immune-related and inflammation-related factors, as well as metabolism like lipid metabolism to some extent. On the other hand, the treatment of Hot herbs against Cold ZHENG CG were mainly related to neurotransmitter and endocrine, which further proved the closer relationship between Cold ZHENG CG and stress-induced factors, and showed therapeutic potential of Hot TCM in this regard. These results preliminarily revealed the mechanisms of action of Cold/Hot herbs in the treatment of CG associated with Cold/Hot ZHENG (Figure 7). These functional characteristics of Cold/Hot herbs in CG formulae for immune regulation and metabolism regulation were also found to be therapeutic targets in researches of other formulae for other diseases [40-43].

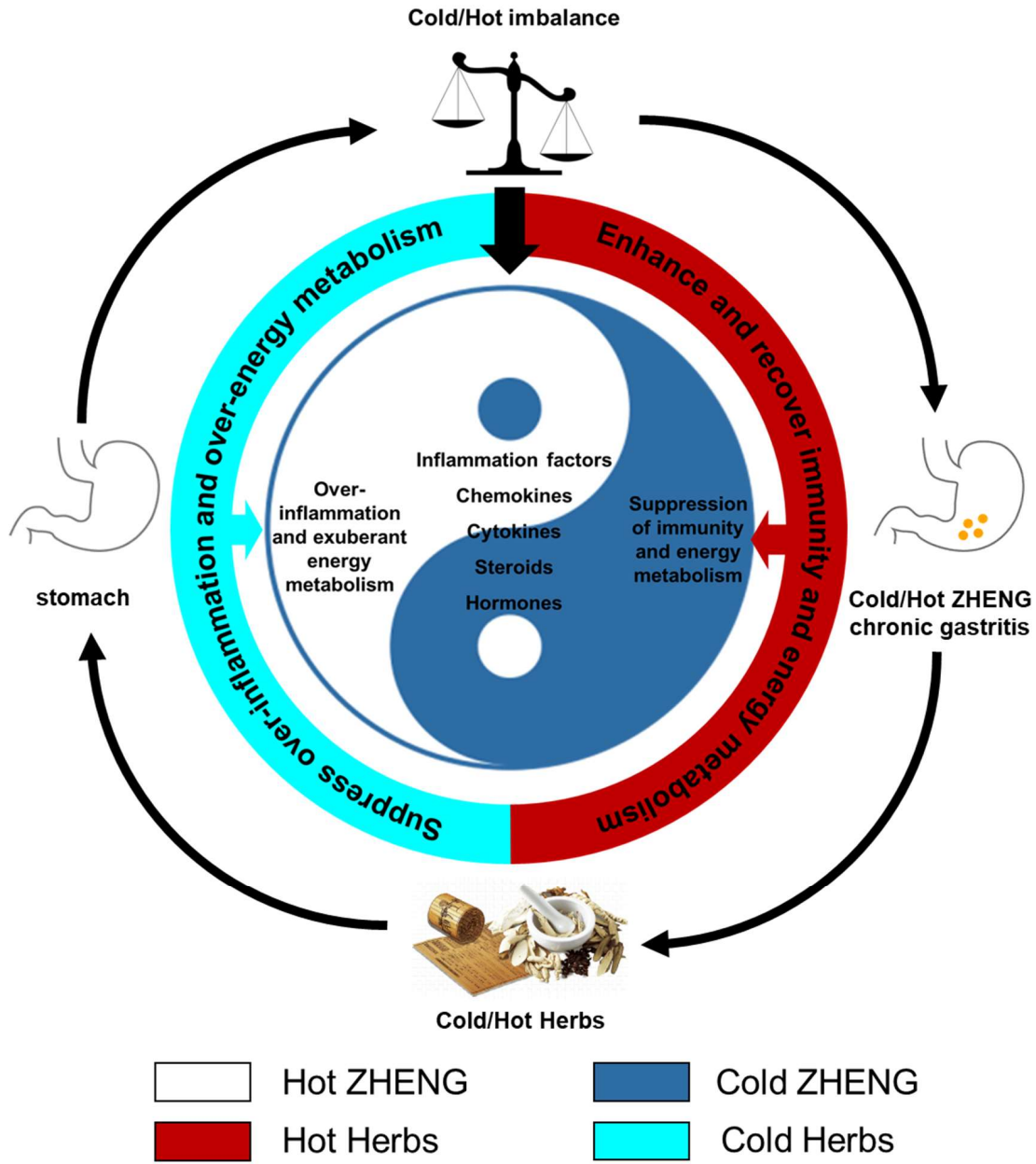

**Figure 7.** Tai Chi Diagram showing the regulation programs of Cold/Hot herbs dominated by bi-directional regulation on inflammation/immune and energy metabolism. Cold herbs showed a trend to suppress the over-inflammation and over exuberant energy metabolism in Hot ZHENG, while Hot herbs preferred to enhance and recover immunity and energy metabolism in Cold ZHENG.

Based on what we found for molecular features of Cold/Hot TCM, two specific networks for the Cold/Hot herbs targets respectively, according to the interaction recoded in STRING database (Figure S3). In Cold herbs targets, INS, TLR2, VEGFA, TNF, IL1R1 and TGFB3 showed a vital role in the targets network of Cold herbs for CG. And in Hot herbs targets, 5-HT related genes HTR1A and HTR1B, NOS1, GABA-related genes GAB family and CRH, as well as somatostatin receptor including SSTR2 and SSTR5 were found in the targets network of Hot herbs for CG. In other words, Hot herbs might regulate Cold ZHENG gastritis through regulating endocrine and energy metabolism. Apart from these genes, targets like CCL2, IL6, JAK2, IL2RA and CXCR1 were also of vital importance in the Hot herbs' targets network, which might infer us that another potential mechanism of Hot herbs against Cold ZEHNG gastritis was to regulate immune responses.

## 3. Materials and Methods

**3.1 Differential analysis and PLS-DA for Cold/Hot ZHENG CG**

To find the significantly differential expressed genes in Cold/Hot ZHENG CG, R package limma [44] were used to construct the generalized linear model. Genes with significant changes ($\log_2$(Fold Change) $\geq 1$ or $\leq -1$, adjust P value < 0.05, BH correction) were considered as differential expressed genes in CAG and CSG, respectively. The DEGs in Cold/Hot ZHENG CG were defined as the DEGs in both Cold/Hot ZHENG CAG and CSG (adjust P value < 0.05, BH correction). PLS-DA (Partial Least Squares Discriminant Analysis) was performed based on R package mixOmics [45] v6.14.1. VIP (Variable Importance in the Projection) for seeds genes of Cold/Hot ZHENG was calculated to estimate the importance of each seeds gene in contributing to distinguish Cold/Hot ZHENG CG (VIP > 1) with function PLSDA.VIP().

**4.2 Enrichment analysis and immune characteristics of Cold/Hot ZHENG CG**

To find the enriched pathways or biological processes which biomolecular features of Cold/Hot ZHENG CG involved in, enrichment analyses were performed based on R package clusterProfiler [46], including KEGG enrichment, GO enrichment and GSEA

(Gene Set Enrichment Analysis). Pathways or biological processes significantly enriched (adjust P value < 0.05, BH correction) were kept for further analysis. Besides, for GSEA results, pathways or biological processes significantly enriched were further divided into 'active' or 'inhibit' based on their positive or negative NES (Normalized Enrichment Score) value.

**3.3 Targets prediction of formulae, herbs and corresponding compounds**

In order to predict the potential targets of compounds in herbs of formulae for CG, based on our network-based computational algorithm, durgCIPHER-CS [14], the genome-wide targets and druggable targets were calculated. Top 100 targets of the druggable targets for each compound were considered as the target profile for each compound. A computational strategy [15] was performed to calculate the holistic targets of our collected formulae for CG and the Cold/Hot herbs that made up them. Targets with significant occurrence (adjust P value < 0.05, BH correction) were listed as the holistic targets of the herbs or formulae.

**3.4 Definition of Cold/Hot herbs' targets and network construction**

Taking the complex composition of herbs into consideration, many biomolecular might be the potential targets for many herbs. A strategy was implemented there to define targets for Hot herbs, Cold herbs and both Cold/Hot herbs. For each target, the counts of Cold/Hot herbs targeting on it were normalized on the total amount of Cold/Hot herbs and the ratio of the normalized counts of Cold/Hot herbs was used to divided targets into three groups, including targets of Hot herbs, targets of Cold herbs and shared targets. For Cold/Hot ZHENG CG, the biological network was constructed by the seeds genes with VIP value larger than 1 and DEGs in both CSG and CAG. Node side of each seeds gene depended on the VIP value. Besides, the biological networks for the targets of Cold/Hot herbs were constructed by the unions of targets of Cold/Hot herbs, respectively and seeds genes with VIP value larger than 1. All the gene-gene interactions in these networks were collected from STRING database [47].

**3.5 Analysis of Cold/Hot TCM and traditional efficacy**

To depict the potential mechanisms of formulae on features of Cold/Hot ZHENG CG, based on the holistic targets of formulae, KEGG and GO enrichment was performed through R package clusterProfiler (adjust P value < 0.05, BH correction). Six labels for traditional effects, including ZI YIN, XIAO JI, SAN HAN, QING RE, HUO XUE and XING QI could paste on these formulae according to TCM experience. And whether the label showed correlation with the composition of Cold/Hot herbs depended on Spearman correlation test (P value < 0.05).

## 4. Discussion

Hot and Cold ZHENG is a dominating theory in TCM and they represent different conditions and phenotypes within one disease [48,49]. For example, in previous studies, significant differential symptoms in a cohort of cases with SARS were closely related to Cold ZHENG [50]. However, the mechanism of Cold/Hot ZHENG, as well as that of "Hot herbs for Cold ZHENG, Cold herbs for Hot ZHENG" remain unclear. Taking two kinds of CG, especially CAG as breakthrough, in this study, we conducted a comprehensive study of Cold/Hot ZHENG CG and corresponding traditional formulae for gastritis composed of Cold/Hot herbs to uncover the mechanism of two traditional properties, Hot and Cold for diseases and herbs originating from ancient China.

Based on microarray datasets for CG of different ZHENG and advanced machine learning algorithms, we found the genes of vital importance as well as pathways or biological processes significantly enriched and constructed a biological network of seeds genes of Cold/Hot ZHENG and DEGs in CG of different ZHENG. It was found that pathways and biological processes related to immune and inflammation regulation were significantly active in Hot ZHENG patients and their related biomolecules were considered as hub nodes in this network due to their high degrees. Meanwhile, pathways and biological processes related to steroid and hormones were found to be active in Cold ZHENG and the corresponding biomolecules were also of vital importance in the network.

In general, the findings in this study could infer us that the main differences between Hot ZHENG and Cold ZHENG CG might be the over-inflammation in Hot

ZHENG, as well as the suppression of immune and energy metabolism in Cold ZHENG. Consistent with our previous study that hormone-related biological processes are predominant in the Cold ZHENG network and immune-related biological processes are predominant in the Hot ZHENG network [11]. More specifically, they might suggest that there are different biological mechanisms between Hot and Cold ZHENG CG, implying that different specific treatment strategies are required for CG depending on ZHENG. Chemokines and cytokines like CCL2 were important biomolecules representing Hot ZHENG from the aspect of immune regulation and inflammation [7], while leptin and nitric oxide involved energy metabolism was another representative difference between Hot and Cold ZHENG.

Then, based on network analysis, we carried out network targets analysis and described the holistic targets profiles of formulae for CG and their corresponding composition of Cold/Hot herbs. Apart from the target information, we also took meridian and Cold/Hot information into account in order to find the potential relationship between Cold/Hot properties and the tissues where TCM might take effect in. Targets of Cold/Hot herbs were divided into three groups and interestingly, it was found that targets of Hot herbs were significantly enriched in metabolism, regulation or endocrine of GABA, hormones, steroids and 5-HT, which were reported to be closely related to energy metabolism and thermogenesis, while targets of Cold herbs were mainly enriched in inflammation regulation like TNF, HIF-1, VEGF signaling pathway as well as immune response including cytokines, chemokines and cellular processes of immune cells. In our constructed biological targets network for Cold/Hot herbs against CG, biomolecules related to inflammation and immune regulation like TNF, TLR2, TGFB3, IL2RG, IL1R1 and VEGF were of vital importance for Cold herbs, while those involved in energy metabolism like SSTR, HTR, GABA and CRH, as well as those related to immune response like CCL2, IL6, IL2RA, IL2RB and JAK2 played important roles in the network for Hot TCM.

Immune-related pathways and biomolecules like TLR2 and CD14 were potential targets of Weifuchun capsule, which was clinically used for CAG, composed of Cold/Hot herbs and performed effect on both Cold/Hot ZHENG, like regulating immune response and anti-inflammation [51]. Huangqi Jianzhong decoction, a formula for Cold ZHENG CG, showed protective effects in CAG rats might be due to the balance of energy expenditure [52]. Taken together, the thermogenesis and immune-enhancing effects of Hot herbs may contribute to the therapeutic effect on Cold ZHENG CG. For herbs with cold property, their treatment of Hot ZHENG CG relies mainly on their anti-inflammatory effects, such as involving the TNF signaling pathway, NF-κB signaling pathway, and VEGF signaling pathway. Zuojin Pill is used in the treatment of Hot CG because it contains the cold property herb (Huanglian, dried rhizome of *Coptis chinensis* Franch., *Coptis teeta* Wall., and *Coptis deltoidea* C.Y.Cheng and P.K.Hsiao), which has been shown to have anti-inflammatory effects on CAG in rats by inhibiting the NF-κB signaling pathway [53]. Another widely used formula for CAG, Moluodan, was reported to reduce the inflammation level, as well as increasing lipid accumulation in MNNG-induced cells [54]. Serum levels of TNF-α, IL-8, and VEGF have been reported to be associated with the severity of CG, the severe degree of neutrophil infiltration in CG, and the severity of precancerous lesions in the stomach, respectively [55]. Weiqi decoction formula has been reported to reduce VEGF levels in CAG rats [56]. The inhibition of the inflammatory response is the main therapeutic route for cold property herbs in the treatment of Hot ZHENG CG. Leptin is considered to be a link between the neuroendocrine and immune systems and could be a possible target for intervention in immunometabolism-mediated pathophysiology and it has been reported that leptin resistance individuals have lower NK cell count and function than normal individuals [57]. Gastric mucosal leptin expression was significantly higher in *H. pylori*-positive patients than in negative patients [58]. Thus, in the context of immune-metabolic imbalances, leptin appears to be the pivotal molecule in the treatment of Cold/Hot CG with Cold/Hot herbs.

Considering the difference in the immune status of Hot and Cold ZHENG CG and the risk of transformation of CG into cancer, we analyzed and compared the association between Cold/Hot ZHENG CG and Cold/Hot tumors. The hot tumors are characterized by immune activation with T cell infiltration, whereas Cold tumors show lack of T cell infiltration or absence [59]. In this study, we found that the biological processes related to immunity, inflammation, cytokines and chemokines were activated in patients with Hot ZHENG CAG and inhibited in patients with Cold ZHENG CAG. Some key seeds genes of Hot/Cold ZHENG play an important role in converting cold tumors into hot tumors through increased T-cell infiltration, such as TGFB, CCL2, VEGF, and TLR [60]. TGFB is an immunosuppressive molecule, and its inhibition increases T cell infiltration [61]. Loss or low expression of specific chemokines and their corresponding receptors, such as CCL2 and CCL5, reduces infiltration of effector T lymphocytes [62]. It has been reported that VEGF could interrupt T-cell priming, inhibit DC maturation and exhaust CTLs [63]. Thus, the recognition of Cold/Hot ZHENG of tumors, such as gastric carcinoma, is particularly important due to the different immune modulation by Cold/Hot herbs treatment.

Last but not least, in order to achieve personalized and precise medical treatment, precision medicine [64], as well as precision TCM [65] which originates from it, provides a new insight for present medical strategy. The diagnosis of Cold/Hot ZHENG is a holistic observation which potentially represents the states of immune regulation and energy metabolism for patients. And decision on Cold/Hot herbs based on corresponding Cold/Hot ZHENG may be a kind of precision TCM on the basis of the macroscopic phenotypes combined with traditional experience. With the help of researches to uncover the mechanism of Cold/Hot ZHENG and herbs, the potential law of precision TCM from the perspective of Cold and Hot may be revealed and thus facilitate newer and more precise medical strategy.

There were still some limitations in this study. Firstly, we summarized the mechanism of Cold/Hot ZHENG CG and Cold/Hot herbs for CG formulae to the level of pathways and biological processes, combined with present studies and proposed some potential biomarkers of these mechanisms. However, these biomarkers haven't been verified in detail. Then, compounds of each TCM in the formulae for CG were collected from databases for TCM. However, the recorded data in these databases might not be as accurate as the result detected by high performance liquid chromatography analysis or other detection analysis. Besides, in the perspective of formulae, we haven't completely uncovered the mechanism for each of them due to their complex composition and some of them were composed of both Hot and Cold TCM, forming complicated effect on chronic gastritis. Fortunately, these shortcomings of data constitution could be partly made up by our network-based algorithms and network targets analysis. And the further analysis of each formulae might need be treated and analyzed respectively to reveal the comprehensive and specific mechanism for them each.

In conclusion, from two starting points, we conducted exhaustive analysis to find vital biomolecules and biological features for Cold/Hot ZHENG CG and Cold/Hot herbs for CG based on the combination of gene expression data, network analysis, statistic models and machine learning algorithms. And it was found that two specific characteristics between Hot and Cold ZHENG were the differences in immune and inflammation responses, as well as those in endocrine, energy metabolism, and thermogenesis. In general, Hot ZHENG CG showed a trend of over-inflammation and exuberant energy metabolism, and that in Cold ZHENG CG was the suppress of immune regulation and energy metabolism. Besides, in the aspect of Cold/Hot herbs, Hot herbs preferred to target on the biomolecules or biological processes of immune response and energy metabolism, while Cold herbs had the potential effect on inflammation and immune regulation. This study didn't only uncover the potential mechanism of Cold/Hot ZHENG CG and Cold/Hot herbs in formulae for CG, but also might potentially provide a new insight for diagnosis of Cold/Hot ZHENG in diseases and further offer better and more precise medication strategies for Cold/Hot ZHENG patients to achieve precision TCM in the treatment of diseases like gastritis.

## Author contributions

Conceptualization, S.L.; supervision, S.L.; investigation, B.Y.W and P.C.; data curation B.Y.W and P.C.; formal analysis B.Y.W.; writing—original draft preparation, B.Y.W. and P.C.; writing—review and editing, S.L. and P.Z.; visualization, B.Y.W. All authors have read and agreed to the published version of the manuscript.

**Declaration of Competing Interest**

The authors declare that they have no known competing financial interests or personal relationships that could have appeared to influence the work reported in this paper.

**Acknowledgments and funding**

This work was supported by the National Natural Science Foundation of China, China [81225025 and 62061160369].

# Reference

1. Du, Y.; Bai, Y.; Xie, P.; Fang, J.; Wang, X.; Hou, X.; Tian, D.; Wang, C.; Liu, Y.; Sha, W.; et al. Chronic gastritis in China: a national multi-center survey. *BMC Gastroenterol* **2014**, *14*, 21, doi:10.1186/1471-230X-14-21.
2. Sipponen, P.; Maaroos, H.I. Chronic gastritis. *Scand J Gastroenterol* **2015**, *50*, 657-667, doi:10.3109/00365521.2015.1019918.
3. Zhang, P.; Yang, M.; Zhang, Y.; Xiao, S.; Lai, X.; Tan, A.; Du, S.; Li, S. Dissecting the Single-Cell Transcriptome Network Underlying Gastric Premalignant Lesions and Early Gastric Cancer. *Cell Rep* **2019**, *27*, 1934-1947 e1935, doi:10.1016/j.celrep.2019.04.052.
4. Qin, F.; Liu, J.Y.; Yuan, J.H. Chaihu-Shugan-San, an oriental herbal preparation, for the treatment of chronic gastritis: a meta-analysis of randomized controlled trials. *J Ethnopharmacol* **2013**, *146*, 433-439, doi:10.1016/j.jep.2013.01.029.
5. Elhadidy, M.G.; El Nashar, E.M.; Alghamdi, M.A.; Samir, S.M. A novel gastroprotective effect of zeaxanthin against stress-induced gastritis in male rats targeting the expression of HIF-1alpha, TFF-1 and MMP-9 through PI3K/Akt/JNK signaling pathway. *Life Sci* **2021**, *273*, 119297, doi:10.1016/j.lfs.2021.119297.
6. Tang, X.D.; Lu, B.; Zhou, L.Y.; Zhan, S.Y.; Li, Z.H.; Li, B.S.; Gao, R.; Wang, F.Y.; Wang, P.; Yang, J.Q.; et al. Clinical practice guideline of Chinese medicine for chronic gastritis. *Chin J Integr Med* **2012**, *18*, 56-71, doi:10.1007/s11655-012-0960-y.
7. Li, R.; Ma, T.; Gu, J.; Liang, X.; Li, S. Imbalanced network biomarkers for traditional Chinese medicine Syndrome in gastritis patients. *Sci Rep* **2013**, *3*, 1543, doi:10.1038/srep01543.
8. Li, S.; Zhang, B. Traditional Chinese medicine network pharmacology: theory, methodology and application. *Chin J Nat Med* **2013**, *11*, 110-120, doi:10.1016/S1875-5364(13)60037-0.
9. Li, S. Network target: a starting point for traditional Chinese medicine network pharmacology. *China Journal of Chinese Materia Medica* **2011**, *36*, 2017-2020.
10. 张彦琼; 李梢. 网络药理学与中医药现代研究的若干进展. *中国药理学与毒理学杂志* **2015**, *29*, 883-892.
11. Li, S.; Zhang, Z.Q.; Wu, L.J.; Zhang, X.G.; Li, Y.D.; Wang, Y.Y. Understanding ZHENG in traditional Chinese medicine in the context of neuro-endocrine-immune network. *IET Syst Biol* **2007**, *1*, 51-60, doi:10.1049/iet-syb:20060032.
12. Zibo Ouyang, S.L. HerbBioMap2.0 Database Platform Building & Mining. Tsinghua university, Tsinghua university, 2014.
13. Wu, Y.; Zhang, F.; Yang, K.; Fang, S.; Bu, D.; Li, H.; Sun, L.; Hu, H.; Gao, K.; Wang, W.; et al. SymMap: an integrative database of traditional Chinese medicine enhanced by symptom mapping. *Nucleic Acids Res* **2019**, *47*, D1110-D1117, doi:10.1093/nar/gky1021.
14. Zhao, S.; Li, S. Network-based relating pharmacological and genomic spaces for drug target identification. *Plos One* **2010**, *5*, e11764, doi:10.1371/journal.pone.0011764.
15. Liang, X.; Li, H.; Li, S. A novel network pharmacology approach to analyse traditional herbal formulae: the Liu-Wei-Di-Huang pill as a case study. *Mol Biosyst* **2014**, *10*, 1014-1022, doi:10.1039/c3mb70507b.
16. Kanehisa, M.; Furumichi, M.; Sato, Y.; Ishiguro-Watanabe, M.; Tanabe, M. KEGG: integrating viruses and cellular organisms. *Nucleic Acids Res* **2021**, *49*, D545-D551, doi:10.1093/nar/gkaa970.

17. Gene Ontology, C. The Gene Ontology resource: enriching a GOld mine. *Nucleic Acids Res* **2021**, *49*, D325-D334, doi:10.1093/nar/gkaa1113.
18. Subramanian, A.; Tamayo, P.; Mootha, V.K.; Mukherjee, S.; Ebert, B.L.; Gillette, M.A.; Paulovich, A.; Pomeroy, S.L.; Golub, T.R.; Lander, E.S.; et al. Gene set enrichment analysis: a knowledge-based approach for interpreting genome-wide expression profiles. *Proc Natl Acad Sci U S A* **2005**, *102*, 15545-15550, doi:10.1073/pnas.0506580102.
19. Huang, Y.; Li, S. Detection of characteristic sub pathway network for angiogenesis based on the comprehensive pathway network. *BMC Bioinformatics* **2010**, *11 Suppl 1*, S32, doi:10.1186/1471-2105-11-S1-S32.
20. Lin, X.M.; Hu, L.; Gu, J.; Wang, R.Y.; Li, L.; Tang, J.; Zhang, B.H.; Yan, X.Z.; Zhu, Y.J.; Hu, C.L.; et al. Choline Kinase alpha Mediates Interactions Between the Epidermal Growth Factor Receptor and Mechanistic Target of Rapamycin Complex 2 in Hepatocellular Carcinoma Cells to Promote Drug Resistance and Xenograft Tumor Progression. *Gastroenterology* **2017**, *152*, 1187-1202, doi:10.1053/j.gastro.2016.12.033.
21. Guo, Y.C.; Bao, C.; Ma, D.C.; Cao, Y.B.; Li, Y.D.; Xie, Z.; Li, S. Network-Based Combinatorial CRISPR-Cas9 Screens Identify Synergistic Modules in Human Cells. *Acs Synth Biol* **2019**, *8*, 482-+, doi:10.1021/acssynbio.8b00237.
22. Newman, A.M.; Steen, C.B.; Liu, C.L.; Gentles, A.J.; Chaudhuri, A.A.; Scherer, F.; Khodadoust, M.S.; Esfahani, M.S.; Luca, B.A.; Steiner, D.; et al. Determining cell type abundance and expression from bulk tissues with digital cytometry. *Nat Biotechnol* **2019**, *37*, 773-782, doi:10.1038/s41587-019-0114-2.
23. Geng, Y.; Wang, J.; Wang, R.; Wang, K.; Xu, Y.; Song, G.; Wu, C.; Yin, Y. Leptin and HER-2 are associated with gastric cancer progression and prognosis of patients. *Biomed Pharmacother* **2012**, *66*, 419-424, doi:10.1016/j.biopha.2012.03.002.
24. Qu, Y.; Wang, X.; Bai, S.; Niu, L.; Zhao, G.; Yao, Y.; Li, B.; Li, H. The effects of TNF-alpha/TNFR2 in regulatory T cells on the microenvironment and progression of gastric cancer. *Int J Cancer* **2022**, *150*, 1373-1391, doi:10.1002/ijc.33873.
25. Zhu, Q.; Zhang, X.; Zhang, L.; Li, W.; Wu, H.; Yuan, X.; Mao, F.; Wang, M.; Zhu, W.; Qian, H.; et al. The IL-6-STAT3 axis mediates a reciprocal crosstalk between cancer-derived mesenchymal stem cells and neutrophils to synergistically prompt gastric cancer progression. *Cell Death Dis* **2014**, *5*, e1295, doi:10.1038/cddis.2014.263.
26. Companioni, O.; Bonet, C.; Garcia, N.; Ramirez-Lazaro, M.J.; Lario, S.; Mendoza, J.; Adrados, M.M.; Poves, E.; Espinosa, L.; Pozo-Kreilinger, J.J.; et al. Genetic variation analysis in a follow-up study of gastric cancer precursor lesions confirms the association of MUC2 variants with the evolution of the lesions and identifies a significant association with NFKB1 and CD14. *Int J Cancer* **2018**, *143*, 2777-2786, doi:10.1002/ijc.31839.
27. Guo, Y.; Bao, C.; Ma, D.; Cao, Y.; Li, Y.; Xie, Z.; Li, S. Network-Based Combinatorial CRISPR-Cas9 Screens Identify Synergistic Modules in Human Cells. *Acs Synth Biol* **2019**, *8*, 482-490, doi:10.1021/acssynbio.8b00237.
28. Guo, Y.; Nie, Q.; MacLean, A.L.; Li, Y.; Lei, J.; Li, S. Multiscale Modeling of Inflammation-Induced Tumorigenesis Reveals Competing Oncogenic and Oncoprotective Roles for Inflammation. *Cancer Res* **2017**, *77*, 6429-6441, doi:10.1158/0008-5472.CAN-17-1662.
29. Wang, L.F.; Liu, Y.S.; Yang, B.; Li, P.; Cheng, X.S.; Xiao, C.X.; Liu, J.J.; Li, S.; Ren, J.L.; Guleng, B. The extracellular matrix protein mindin attenuates colon cancer progression by blocking angiogenesis via Egr-1-mediated regulation. *Oncogene* **2018**, *37*, 601-615, doi:10.1038/onc.2017.359.

30. Zhao, X.; Fu, J.; Xu, A.; Yu, L.; Zhu, J.; Dai, R.; Su, B.; Luo, T.; Li, N.; Qin, W.; et al. Gankyrin drives malignant transformation of chronic liver damage-mediated fibrosis via the Rac1/JNK pathway. *Cell Death Dis* **2015**, *6*, e1751, doi:10.1038/cddis.2015.120.

31. Lei, Y.; Li, S.; Liu, Z.; Wan, F.; Tian, T.; Li, S.; Zhao, D.; Zeng, J. A deep-learning framework for multi-level peptide-protein interaction prediction. *Nat Commun* **2021**, *12*, 5465, doi:10.1038/s41467-021-25772-4.

32. Park, H.K.; Ahima, R.S. Physiology of leptin: energy homeostasis, neuroendocrine function and metabolism. *Metabolism* **2015**, *64*, 24-34, doi:10.1016/j.metabol.2014.08.004.

33. Lee, B.; Kim, J.; An, T.; Kim, S.; Patel, E.M.; Raber, J.; Lee, S.K.; Lee, S.; Lee, J.W. Dlx1/2 and Otp coordinate the production of hypothalamic GHRH- and AgRP-neurons. *Nat Commun* **2018**, *9*, 2026, doi:10.1038/s41467-018-04377-4.

34. Atrens, D.M.; Menendez, J.A. Somatostatin and the paraventricular hypothalamus: modulation of energy balance. *Brain Res* **1993**, *630*, 238-244, doi:10.1016/0006-8993(93)90662-7.

35. Oh, C.M.; Namkung, J.; Go, Y.; Shong, K.E.; Kim, K.; Kim, H.; Park, B.Y.; Lee, H.W.; Jeon, Y.H.; Song, J.; et al. Regulation of systemic energy homeostasis by serotonin in adipose tissues. *Nat Commun* **2015**, *6*, 6794, doi:10.1038/ncomms7794.

36. Richard, D. Cognitive and autonomic determinants of energy homeostasis in obesity. *Nat Rev Endocrinol* **2015**, *11*, 489-501, doi:10.1038/nrendo.2015.103.

37. Adamovsky, O.; Buerger, A.N.; Vespalcova, H.; Sohag, S.R.; Hanlon, A.T.; Ginn, P.E.; Craft, S.L.; Smatana, S.; Budinska, E.; Persico, M.; et al. Evaluation of Microbiome-Host Relationships in the Zebrafish Gastrointestinal System Reveals Adaptive Immunity Is a Target of Bis(2-ethylhexyl) Phthalate (DEHP) Exposure. *Environ Sci Technol* **2020**, *54*, 5719-5728, doi:10.1021/acs.est.0c00628.

38. Tian, W.; Zhang, N.; Jin, R.; Feng, Y.; Wang, S.; Gao, S.; Gao, R.; Wu, G.; Tian, D.; Tan, W.; et al. Immune suppression in the early stage of COVID-19 disease. *Nat Commun* **2020**, *11*, 5859, doi:10.1038/s41467-020-19706-9.

39. Kim, S.; Chen, J.; Cheng, T.; Gindulyte, A.; He, J.; He, S.; Li, Q.; Shoemaker, B.A.; Thiessen, P.A.; Yu, B.; et al. PubChem in 2021: new data content and improved web interfaces. *Nucleic Acids Res* **2021**, *49*, D1388-D1395, doi:10.1093/nar/gkaa971.

40. Zhang, S.; Lai, X.; Wang, X.; Liu, G.; Wang, Z.; Cao, L.; Zhang, X.; Xiao, W.; Li, S. Deciphering the Pharmacological Mechanisms of Guizhi-Fuling Capsule on Primary Dysmenorrhea Through Network Pharmacology. *Front Pharmacol* **2021**, *12*, 613104, doi:10.3389/fphar.2021.613104.

41. Li, S.; Lu, A.P.; Wang, Y.Y.; Li, Y.D. Suppressive effects of a Chinese herbal medicine qing-luo-yin extract on the angiogenesis of collagen-induced arthritis in rats. *Am J Chin Med* **2003**, *31*, 713-720, doi:10.1142/S0192415X03001430.

42. Zhou, W.; Lai, X.; Wang, X.; Yao, X.; Wang, W.; Li, S. Network pharmacology to explore the anti-inflammatory mechanism of Xuebijing in the treatment of sepsis. *Phytomedicine* **2021**, *85*, 153543, doi:10.1016/j.phymed.2021.153543.

43. Zuo, J.; Wang, X.; Liu, Y.; Ye, J.; Liu, Q.; Li, Y.; Li, S. Integrating Network Pharmacology and Metabolomics Study on Anti-rheumatic Mechanisms and Antagonistic Effects Against Methotrexate-Induced Toxicity of Qing-Luo-Yin. *Front Pharmacol* **2018**, *9*, 1472, doi:10.3389/fphar.2018.01472.
44. Ritchie, M.E.; Phipson, B.; Wu, D.; Hu, Y.; Law, C.W.; Shi, W.; Smyth, G.K. limma powers differential expression analyses for RNA-sequencing and microarray studies. *Nucleic Acids Res* **2015**, *43*, e47, doi:10.1093/nar/gkv007.
45. Rohart, F.; Gautier, B.; Singh, A.; Le Cao, K.A. mixOmics: An R package for 'omics feature selection and multiple data integration. *PLoS Comput Biol* **2017**, *13*, e1005752, doi:10.1371/journal.pcbi.1005752.
46. Yu, G.; Wang, L.G.; Han, Y.; He, Q.Y. clusterProfiler: an R package for comparing biological themes among gene clusters. *Omics* **2012**, *16*, 284-287, doi:10.1089/omi.2011.0118.
47. Szklarczyk, D.; Gable, A.L.; Nastou, K.C.; Lyon, D.; Kirsch, R.; Pyysalo, S.; Doncheva, N.T.; Legeay, M.; Fang, T.; Bork, P.; et al. The STRING database in 2021: customizable protein-protein networks, and functional characterization of user-uploaded gene/measurement sets. *Nucleic Acids Res* **2021**, *49*, D605-D612, doi:10.1093/nar/gkaa1074.
48. 吕爱平; 李梢; 王永炎. 从主观症状的客观规律探索中医证候分类的科学基础. *中医杂志* **2005**, *01*, 4-6.
49. Su, S.B.; Jia, W.; Lu, A.P.; Li, S. Evidence-Based ZHENG: A Traditional Chinese Medicine Syndrome 2013. *Evid-Based Compl Alt* **2014**, *2014*, doi:Artn 484201
10.1155/2014/484201.
50. Li, S.; Wang, R.Q.; Zhang, Y.L.; Zhang, X.G.; Layon, A.J.; Li, Y.D.; Chen, M.Z. Symptom combinations associated with outcome and therapeutic effects in a cohort of cases with SARS. *Am J Chinese Med* **2006**, *34*, 937-947, doi:Doi 10.1142/S0192415x06004417.
51. Wang, B.; Zhou, W.; Zhang, H.; Wang, W.; Zhang, B.; Li, S. Exploring the effect of Weifuchun capsule on the toll-like receptor pathway mediated HES6 and immune regulation against chronic atrophic gastritis. *J Ethnopharmacol* **2022**, *303*, 115930, doi:10.1016/j.jep.2022.115930.
52. Liu, Y.; Jin, Z.; Qin, X.; Zheng, Q. Urinary metabolomics research for Huangqi Jianzhong Tang against chronic atrophic gastritis rats based on (1) H NMR and UPLC-Q/TOF MS. *J Pharm Pharmacol* **2020**, *72*, 748-760, doi:10.1111/jphp.13242.
53. Wen, J.; Wu, S.; Ma, X.; Zhao, Y. Zuojin Pill attenuates Helicobacter pylori-induced chronic atrophic gastritis in rats and improves gastric epithelial cells function in GES-1 cells. *J Ethnopharmacol* **2022**, *285*, 114855, doi:10.1016/j.jep.2021.114855.
54. Zhou, W.; Zhang, H.; Wang, X.; Kang, J.; Guo, W.; Zhou, L.; Liu, H.; Wang, M.; Jia, R.; Du, X.; et al. Network pharmacology to unveil the mechanism of Moluodan in the treatment of chronic atrophic gastritis. *Phytomedicine* **2022**, *95*, 153837, doi:10.1016/j.phymed.2021.153837.
55. Siregar, G.A.; Halim, S.; Sitepu, V.R. Serum TNF-a, IL-8, VEGF levels in Helicobacter pylori infection and their association with degree of gastritis. *Acta Med Indones* **2015**, *47*, 120-126.
56. Yin, J.; Yi, J.; Yang, C.; Xu, B.; Lin, J.; Hu, H.; Wu, X.; Shi, H.; Fei, X. Weiqi Decoction Attenuated Chronic Atrophic Gastritis with Precancerous Lesion through Regulating Microcirculation Disturbance and HIF-1alpha Signaling Pathway. *Evid Based Complement Alternat Med* **2019**, *2019*, 2651037, doi:10.1155/2019/2651037.
57. Abella, V.; Scotece, M.; Conde, J.; Pino, J.; Gonzalez-Gay, M.A.; Gomez-Reino, J.J.; Mera,

A.; Lago, F.; Gomez, R.; Gualillo, O. Leptin in the interplay of inflammation, metabolism and immune system disorders. *Nat Rev Rheumatol* **2017**, *13*, 100-109, doi:10.1038/nrrheum.2016.209.

58. Jun, D.W.; Lee, O.Y.; Lee, Y.Y.; Choi, H.S.; Kim, T.H.; Yoon, B.C. Correlation between gastrointestinal symptoms and gastric leptin and ghrelin expression in patients with gastritis. *Dig Dis Sci* **2007**, *52*, 2866-2872, doi:10.1007/s10620-006-9651-x.
59. Duan, Q.; Zhang, H.; Zheng, J.; Zhang, L. Turning Cold into Hot: Firing up the Tumor Microenvironment. *Trends Cancer* **2020**, *6*, 605-618, doi:10.1016/j.trecan.2020.02.022.
60. Liu, Y.T.; Sun, Z.J. Turning cold tumors into hot tumors by improving T-cell infiltration. *Theranostics* **2021**, *11*, 5365-5386, doi:10.7150/thno.58390.
61. Wang, M.; Wang, S.; Desai, J.; Trapani, J.A.; Neeson, P.J. Therapeutic strategies to remodel immunologically cold tumors. *Clin Transl Immunology* **2020**, *9*, e1226, doi:10.1002/cti2.1226.
62. DeNardo, D.G.; Ruffell, B. Macrophages as regulators of tumour immunity and immunotherapy. *Nat Rev Immunol* **2019**, *19*, 369-382, doi:10.1038/s41577-019-0127-6.
63. Ni, J.J.; Zhang, Z.Z.; Ge, M.J.; Chen, J.Y.; Zhuo, W. Immune-based combination therapy to convert immunologically cold tumors into hot tumors: an update and new insights. *Acta Pharmacol Sin* **2022**, doi:10.1038/s41401-022-00953-z.
64. Aronson, S.J.; Rehm, H.L. Building the foundation for genomics in precision medicine. *Nature* **2015**, *526*, 336-342, doi:10.1038/nature15816.
65. Wang, W.J.; Zhang, T. Integration of traditional Chinese medicine and Western medicine in the era of precision medicine. *J Integr Med* **2017**, *15*, 1-7, doi:10.1016/S2095-4964(17)60314-5.